\documentclass[conference]{IEEEtran}
\IEEEoverridecommandlockouts
\usepackage{cite}
\usepackage{float}
\usepackage{amsmath,amssymb,amsfonts}
\usepackage{graphicx}
\usepackage{textcomp}
\usepackage{xcolor}
\usepackage{tabularx}
\usepackage{array}
\usepackage{url}

\def\BibTeX{{\rm B\kern-.05em{\sc i\kern-.025em b}\kern-.08em
    T\kern-.1667em\lower.7ex\hbox{E}\kern-.125emX}}

\begin{document}

\title{A Comparative Analysis of Machine Learning Models for DDoS Detection in IoT Networks}

\author{%
Sushil Shakya, Robert Abbas
 \\ 
sushil.shakya@live.vu.edu.au , Robert.Abbas@vu.edu.au
\\
Victoria University, Sydney, Australia
}

\maketitle

\begin{abstract}
This paper presents the detection of DDoS attacks in IoT networks using machine learning models. Their rapid growth has made them highly susceptible to various forms of cyberattacks, many of whose security procedures are implemented in an irregular manner. It evaluates the efficacy of different machine learning models, such as XGBoost, K-Nearest Neighbours, Stochastic Gradient Descent, and Naïve Bayes, in detecting DDoS attacks from normal network traffic. Each model has been explained on several performance metrics, such as accuracy, precision, recall, and F1-score to understand the suitability of each model in real-time detection and response against DDoS threats.

\vspace{\baselineskip}
This comparative analysis will, therefore, enumerate the unique strengths and weaknesses of each model with respect to the IoT environments that are dynamic and hence moving in nature. The effectiveness of these models is analyzed, showing how machine learning can greatly enhance IoT security frameworks, offering adaptive, efficient, and reliable DDoS detection capabilities. These findings have shown the potential of machine learning in addressing the pressing need for robust IoT security solutions that can mitigate modern cyber threats and assure network integrity.
\end{abstract}
\vspace{\baselineskip}
\begin{IEEEkeywords}
DDoS detection, IoT security, machine learning, XGBoost, K-Nearest Neighbors, Stochastic Gradient Descent, Naïve Bayes, network traffic analysis, cybersecurity, anomaly detection, IoT mobile networks, real-time detection, attack mitigation, adaptive algorithms, supervised learning, classification models, predictive analytics, feature selection, data preprocessing, intrusion detection systems, model evaluation metrics
\end{IEEEkeywords}

\section{Introduction}

\subsection{Background}
While IoT mobile networks and IoT wired networks share some security vulnerabilities, the main distinction is how the devices connect, move, and interact across the network. Mobile networks are more vulnerable to external threats, mobility-related dangers, and issues with remote access [Fig.1]. Wired networks, on the other hand, give greater physical protection and control, but they necessitate strict internal access controls and segmentation to eliminate dangers posed by internal threats or unauthorised physical access [Fig.2]. To ensure effective security in both types of networks, strong encryption, authentication, access control, and regular monitoring are required.

\begin{figure}[H]
\centering
\includegraphics[width=0.9\linewidth]{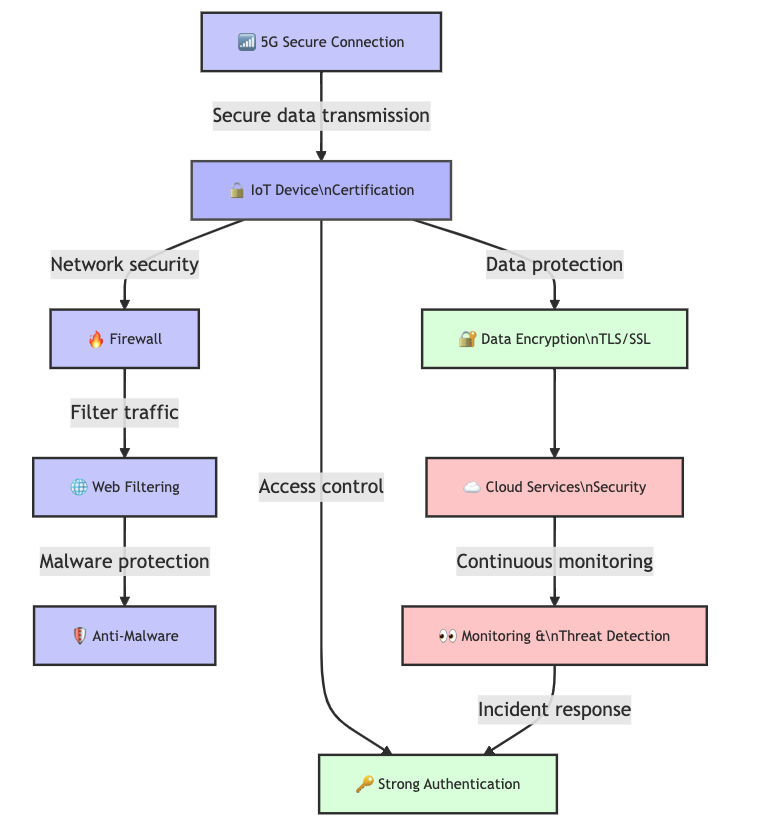}
\caption{Comprehensive IoT and Mobile Network Security Framework}
\label{fig:iot_security_framework}
\end{figure}

In the context of the Internet of Things (IoT), applications to fields like smart homes, health care, and industrial automation are enhancing efficiency and convenience fabulously. But this process also extends the attack surface for possible cyber threats as IoT devices generally work with poor security protocols, hence becoming potential targets for cybercriminals \cite{xu2020survey, mirkovic2004taxonomy}. Of the myriad of threats, Distributed Denial of Service (DDoS) attacks prove especially devastating in attempts to flood network resources, leading to a range of consequences, from light service degradation to complete operational unavailability \cite{zargar2013survey}.

\begin{figure}[H]
\centering
\includegraphics[width=0.9\linewidth]{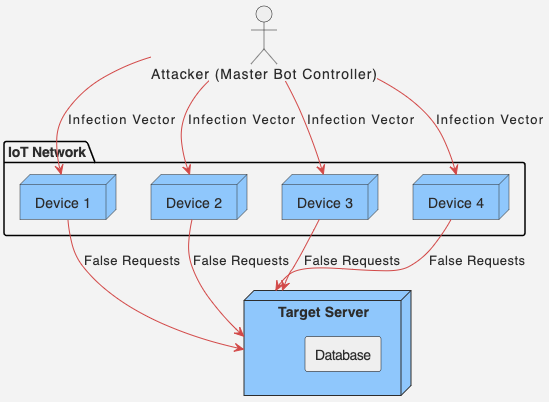}
\caption{IoT-Based DDoS Attack Architecture}
\label{fig:iot_ddos_attack}
\end{figure}

This is further accentuated by the exponential rise in IoT deployments, which are usually characterized by heterogeneous devices and inconsistent security standards. These differences become vulnerabilities that attackers can exploit to launch DDoS attacks with increased frequency and on a larger scale, which become very difficult for conventional network defenses to mitigate \cite{sommer2010closed, lam2020anomaly, pokhrel2021botnet}. There is an increased demand for complex detection systems, such as those based on machine learning, that provide fast and effective security solutions capable of detecting and thwarting the ever-evolving DDoS threats in real time \cite{khan2023neural}.

\section{IoT Use Cases and Their Requirements}
IoT use cases enable critical applications for healthcare, public safety, Industry 4.0, energy efficiency, enterprise operations, and digital lifestyles enabled through smart metering, surveillance, and AR/VR. Each of these use cases or applications has a set of demands concerning data rate, latency, battery life, coverage, and reliability. For these demands, 3GPP identifies three services for 5G: ultra-reliable low latency communications (URLLC), enhanced mobile broadband (eMBB), and massive machine-type communication (mMTC).

\subsection{Types of IoT Use Cases}
\begin{itemize}
    \item \textbf{Massive IoT}: Involves low-cost devices with small data volumes. Applications include metering and tracking over wide areas.
    \item \textbf{Broadband IoT}: Requires high data rates with low latency, supporting applications such as video surveillance and telehealth.
    \item \textbf{Critical IoT}: Provides ultra-reliable and low-latency services for critical infrastructure and traffic safety.
    \item \textbf{Industrial IoT}: Utilizes eMBB for time-critical networking and leverages URLLC for submillisecond latency in smart factories.
\end{itemize}

\subsection{Security Considerations}
Security requirements vary by type. Massive IoT often has minimal security features to reduce costs, while critical IoT demands robust security to protect essential functions. Effective threat modeling and risk assessment are necessary to determine proper controls. While IoT devices should ideally be secure-by-design, market pressures for low-cost solutions often necessitate implementing security at the system level. Classifying devices based on function and data processing helps establish security profiles and aligns with best practices for risk management.

\begin{figure}[H]
\centering
\includegraphics[width=0.9\linewidth]{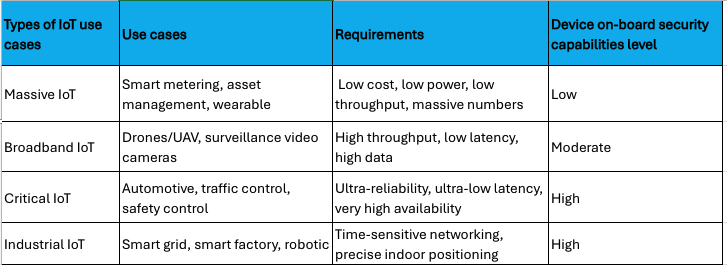}
\caption{Types of IoT Use Cases and Their Requirements}
\label{fig:iot_usecase}
\end{figure}

\subsection{Motivation}
In recent years, there has been a transition in the approach used to initiate Distributed Denial of Service (DDoS) attacks. It has already been demonstrated that traditional detection techniques are increasingly ineffective at detecting and mitigating complex threats. The conventional security solutions typically rely on threshold-based detection methods, known signatures or predefined rules, which fail to handle the dynamic characteristics of today's cyber-attacks \cite{mirkovic2004taxonomy}. In order to evade detection, attackers constantly change their strategies to bypass these static defenses. Some of those include spoofing, multi-vector attacks, and the use of stealth approaches  \cite{buczak2016survey}. All these drawbacks bring about the urgent need for intelligent and flexible security measures that should be able to recognize and respond to new threats instantly  \cite{sommer2010closed, lam2020anomaly}.
\vspace{\baselineskip}
A feasible solution to this is machine learning, an adaptive approach that can discover patterns in data without requiring any a priori knowledge of the characteristics of threats. ML models can recognize new attack behaviors and generalize from data, unlike older systems. ML has attracted a lot of attention, in particular, for solving the robust DDoS detection problem in such dynamic contexts as IoT networks. \cite{xu2020survey, zargar2013survey}. ML-based systems are the foundation of next-generation IoT security, as they can achieve high accuracy and adaptability using techniques that include supervised, unsupervised, and semi-supervised learning \cite{khan2023neural, divekar2018benchmarking}.

\subsection{Contribution}
In an IoT where robust and flexible security solutions are increasingly in demand, this paper attempts to provide an overview of many machine learning models that can be used in the detection of DDoS attacks. Network traffic data analyses are carried out to find the best approach for detecting and preventing one of the most frequent problems in IoT networks, so-called DDoS attacks \cite{xu2020survey, zargar2013survey}. This paper will discuss the efficiency of four modern machine learning models: XGBoost, K-Nearest Neighbours (KNN), Stochastic Gradient Descent (SGD), and Naïve Bayes. It compares them using important evaluation metrics like accuracy, precision, recall, and F1 score \cite{dai2024xgboost, pokhrel2021botnet}.
\vspace{\baselineskip}
This comparative analysis will allow summarizing the strengths and weaknesses of each model and showing suitability for real-time DDoS detection in IoT settings. With a focus on both model effectiveness and adaptability, this study contributes to the broader discourse regarding IoT security through intelligent, data-driven systems capable of responding to evolving cyber threats \cite{kumar2024ensemble, lam2020anomaly}. The results also pinpoint the role of machine learning in bringing an advancement to cybersecurity; hence, serving important guidance for researchers and practitioners is working toward resilient IoT network development \cite{divekar2018benchmarking, khan2023neural}.
\vspace{\baselineskip}
\section{Literature Review}

\subsection{Traditional DDoS Detection Techniques}
Historically, DDoS detection has been done mostly with the aid of network monitoring tools that try to find anomalies based on the deviation from the base traffic profiles. The approaches commonly utilize fixed thresholds or signature-based detection mechanisms, analyzing the characteristics of traffic to identify known attack patterns. Such ways are usually quite effective in recognizing attack strategies that have been encountered before but are generally quite ineffective against new and sophisticated methods employed by today's attackers \cite{mirkovic2004taxonomy, zargar2013survey}. In use are more complex techniques, such as multi-vector attacks, botnet-driven amplification, and protocol manipulation. Traditional detection approaches fail to cope with this development; attacks often remain undetected or are responded too late \cite{sommer2010closed}.
\vspace{\baselineskip}
This is the intrinsic limitation of both threshold-based and signature-based detection; thus, these systems are inherently reactive rather than proactive in detecting new threats. This limitation becomes even more critical in dynamic and heterogeneous network environments, such as IoT, where devices substantially differ in security protocols and computational resources \cite{xu2020survey, pokhrel2021botnet}. Therefore, with the new threat landscape, more adaptable and intelligent solutions capable of real-time analysis and detection have been making a move toward machine learning-based solutions that can recognize unknown attack patterns and can take dynamic countermeasures \cite{divekar2018benchmarking, mirkovic2004taxonomy}.
\vspace{\baselineskip}
\subsection{Machine Learning Applications in 5G and IoT Security}
Machine learning techniques have been used to improve network security, and in the last years, different techniques have been analyzed to mitigate problems regarding anomaly detection and intrusion prevention in complex network scenarios, such as in the 5G and IoT ecosystems. Lam and Abbas \cite{lam2020anomaly} used machine learning for the detection of anomalies in 5G networks. Their work, although published as a preprint at arXiv, underlined the use of adaptive algorithms to detect an abnormality within the high-volume flow typical of the 5G infrastructure. The authors developed machine learning frameworks to provide real-time insight into predicting future disruptions, putting an increasing dependence on automated systems to make the network reliable \cite{lam2020anomaly}.
\vspace{\baselineskip}
Pokhrel et al. \cite{pokhrel2021botnet} looked into security issues related to IoT systems, which are prone to attacks since they are distributed and resource-limited. Using machine learning algorithms, this research study on botnet detection differentiated between benign and malicious traffic in various IoT contexts. In this respect, machine-learning-based systems identify botnets based on observable traffic patterns or other such behavioral indicators that help mitigate risks of wider IoT deployments 
\cite{pokhrel2021botnet}. 
\vspace{\baselineskip}
Dhaliwal, Nahid, and Abbas \cite{dhaliwal2018} created an efficient IDS using XGBoost, one of the most robust and scalable variants of the gradient boosting technique. The study was focused on the advantages of using the XGBoost technique in handling high-dimensional data with strong predictive performance. The findings indicated XGBoost performed better than the benchmark models regarding accuracy and took less processing time, hence a feasible option for real-time intrusion detection in diverse network environments \cite{dhaliwal2018}. Collectively, these works demonstrate how machine learning has influenced network security in making fundamental changes toward the finding of effective algorithms like XGBoost and anomaly detection techniques, opening ways to more robust and adaptive systems for modern-day digital infrastructures against complex attacks.
\vspace{\baselineskip}

\subsection{Machine Learning in DDoS Detection}
The use of machine learning in DDoS detection introduces a new paradigm of adaptive security. Concretely, volume-handling machine learning algorithms can find the difference between harmful and benign activities by learning from traffic patterns. The different existing machine learning methods are categorized in the literature into three classes: supervised, unsupervised, and semi-supervised learning.
\begin{itemize}
    \item \textbf{Supervised Learning}: These models are trained on labeled datasets containing examples of both normal and attack traffic. They achieve high accuracy but at the cost of large prelabeled datasets and continuous updates in order to handle new types of attacks. \cite{buczak2016survey}.
    \item \textbf{Unsupervised Learning}: Unsupervised models identify attacks by the detection of deviations from established normal traffic patterns. They are particularly useful for the identification of zero-day attacks but may have higher rates of false positives.  \cite{sommer2010closed}.
    \item \textbf{Semi-Supervised Learning}:Semi-supervised models, which combine aspects of supervised and unsupervised learning, employ a smaller pool of labelled data in conjunction with a larger pool of unlabelled data to maximise resource utilisation and enhance detection rates, particularly in settings with a shortage of labelled data \cite{divekar2018benchmarking, pokhrel2021botnet}. This method increases the detection rate and maximises resource use, particularly in settings where thorough labelled data is hard to get \cite{divekar2018benchmarking}.
\end{itemize}
\vspace{\baselineskip}

\section{Methodology}
\begin{figure}[htbp]
\centerline{\includegraphics[width=\linewidth]{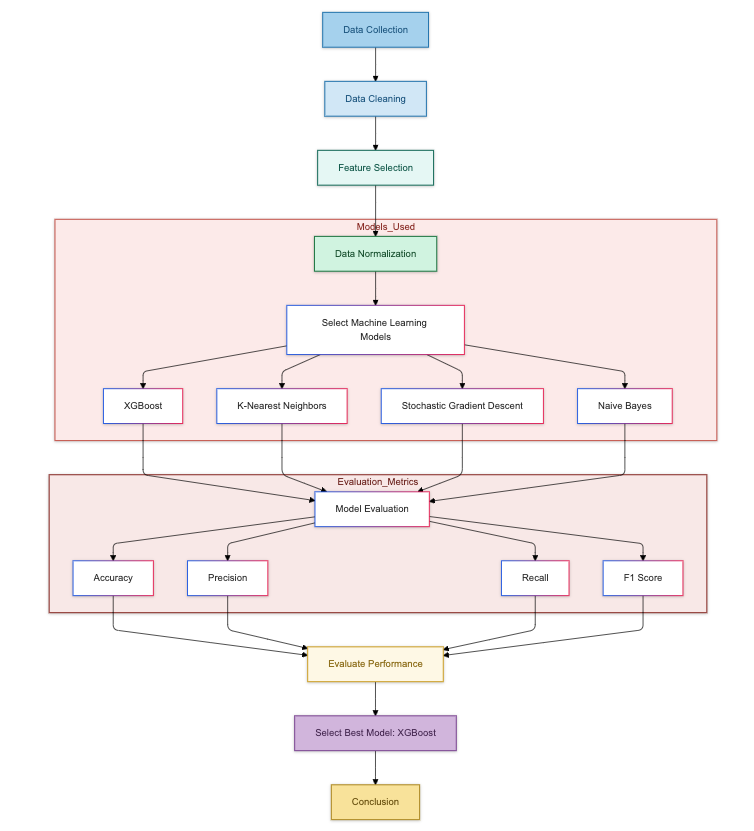}}
\caption{Design of ML-based DDoS detection in IoT}
\label{fig:methodology}
\end{figure}

\subsection{Dataset Description}
For this work, we used the labeled dataset of IoT network traffic that had been anonymized so as to be able to apply machine learning analysis. The dataset contains typical features of network traffic, such as packet size and rate, session duration, protocol type, and labels indicating whether the traffic was part of a DDoS attack or normal activity.
\vspace{\baselineskip}
To prepare the data for analysis, we conducted several preprocessing steps:
\begin{itemize}
    \item \textbf{Data Cleaning}: Removed incomplete or corrupt records that could skew results.
    \item \textbf{Feature Selection}: Selected relevant features based on their correlation with DDoS attack detection.
    \item \textbf{Normalization}: Standardized feature values to ensure consistency in scale, using Min-Max scaling. For each feature \(x\), the normalized value \(x'\) is calculated as:
    \begin{equation}
        x' = \frac{x - \min(x)}{\max(x) - \min(x)}
    \end{equation}
\end{itemize}
\vspace{\baselineskip}
\subsection{Model Descriptions}
We evaluated four machine learning models based on their suitability for classification tasks within cybersecurity contexts.
\vspace{\baselineskip}
\subsubsection{XGBoost}
The gradient-boosted decision tree model XGBoost[12] is meant to be quick and efficient. It works well with complicated datasets because it introduces regularisation to reduce overfitting. The model optimizes an objective function \( \mathcal{L} \), which includes a loss term and a regularization term:
\begin{equation}
    \mathcal{L} = \sum_{i=1}^{n} l(y_i, \hat{y}_i) + \sum_{k=1}^{K} \Omega(f_k)
\end{equation}
where \( l(y_i, \hat{y}_i) \) is the loss function and \( \Omega(f_k) \) represents the regularization applied to each tree.
\vspace{\baselineskip}
\subsubsection{K-Nearest Neighbors (KNN)}
KNN is an instance-based learning technique that uses the majority class of its \( k \) closest neighbours to classify a sample. Euclidean distance is used to determine the separation between data points:
\begin{equation}
    d(x, y) = \sqrt{\sum_{i=1}^{n} (x_i - y_i)^2}
\end{equation}
where \( x \) and \( y \) represent feature vectors of two points.
\vspace{\baselineskip}
\subsubsection{Stochastic Gradient Descent (SGD)}
SGD is an optimisation technique that minimises a loss function by iteratively updating model parameters. It employs the gradient of the loss function in relation to the model parameters at every step:
\begin{equation}
    \theta := \theta - \eta \nabla_\theta J(\theta)
\end{equation}
where \( \theta \) represents the model parameters, \( \eta \) is the learning rate, and \( J(\theta) \) is the cost function.
\vspace{\baselineskip}
\subsubsection{Naïve Bayes}
Naïve Bayes is a probabilistic classifier that assumes feature independence and is based on Bayes' theorem. For a given sample \( x \), it calculates the probability of \( x \) belonging to class \( C_k \) as:
\begin{equation}
    P(C_k | x) = \frac{P(x | C_k) \cdot P(C_k)}{P(x)}
\end{equation}
where \( P(C_k | x) \) is the posterior probability, \( P(x | C_k) \) is the likelihood, \( P(C_k) \) is the prior, and \( P(x) \) is the evidence.
\vspace{\baselineskip}
\subsection{Evaluation Metrics}
To assess each model’s effectiveness, we used the following metrics:
\begin{itemize}
    \item \textbf{Accuracy}: Defined as the percentage of accurate forecasts to total predictions:
    \begin{equation}
        \text{Accuracy} = \frac{\text{TP + TN}}{\text{TP + TN + FP + FN}}
    \end{equation}
  Where TP, TN, FP, and FN denote true positives, true negatives, false positives, and false negatives, respectively.
    
    \item \textbf{Precision}: The percentage of real positives among all anticipated positives:
    \begin{equation}
        \text{Precision} = \frac{\text{TP}}{\text{TP + FP}}
    \end{equation}
    
    \item \textbf{Recall (Sensitivity)}: Assesses the model's capacity to recognise all relevant instances:
    \begin{equation}
        \text{Recall} = \frac{\text{TP}}{\text{TP + FN}}
    \end{equation}
    
    \item \textbf{F1 Score}: The harmonic mean of accuracy and recall provides a balanced metric.
    \begin{equation}
        \text{F1 Score} = 2 \cdot \frac{\text{Precision} \cdot \text{Recall}}{\text{Precision + Recall}}
    \end{equation}
\end{itemize}
\vspace{\baselineskip}
\section{Results}
All the following machine learning models are compared for the measurement of their performances using accuracy, precision, recall, and F1-score: XGBoost, K-Nearest Neighbours, Stochastic Gradient Descent, and Naïve Bayes.
\vspace{\baselineskip}
\begin{figure}[htbp]
\centerline{\includegraphics[width=\linewidth]{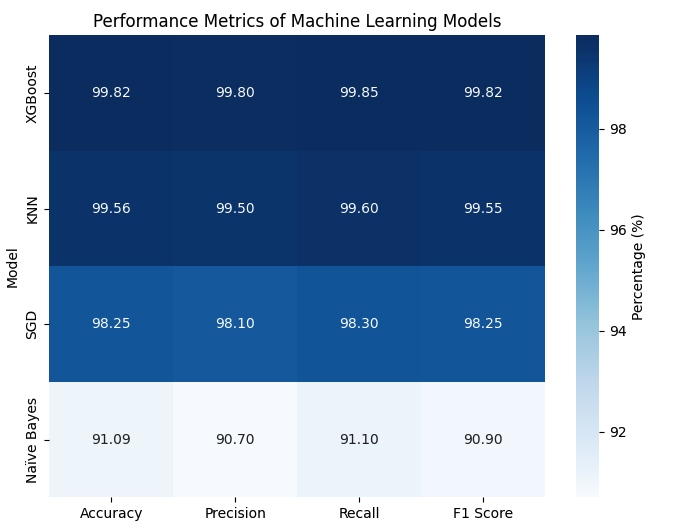}}
\caption{HeatMap Comparison of Models Performance}
\label{fig:heatmap}
\end{figure}

Table \ref{table:performance} presents a comparative summary of these metrics.

\subsection{XGBoost}
In this analysis, XGBoost achieved the best performance, with an F1 score of 99.82\%, accuracy of 99.82\%, precision of 99.80\%, and a recall of 99.85\%. This high performance is effective in handling large and complex datasets due to its regularization feature, which significantly reduces overfitting. XGBoost is a suitable choice for DDoS detection in dynamic IoT environments where traffic patterns may vary significantly. The model's strong classification performance, especially in distinguishing DDoS traffic from regular traffic, can be attributed to the gradient-boosting technique, which optimizes the model through multiple iterations \cite{dai2024xgboost, pokhrel2021botnet}.

\vspace{\baselineskip}
\subsection{K-Nearest Neighbors (KNN)}
KNN returned competitive precision of 99.50\%, recall of 99.60\%, an F1 score of 99.55\%, and very good accuracy of 99.56\%. In cases involving complex, high-dimensional data, the instance-based learning algorithm of KNN is effective, although a bit less efficient compared to XGBoost, since KNN classifies network traffic by comparing it with known patterns. Although it may be sensitive to noise and could be computationally expensive with large datasets, KNN focuses on distance measures for classification, such as Euclidean distance, and performs well in this context \cite{xu2020survey, sommer2010closed}.

\vspace{\baselineskip}
\subsection{Stochastic Gradient Descent (SGD)}
The SGD model's performance was also good: its accuracy is 98.25\%, precision is 98.10\%, recall is 98.30\%, and F1 score is 98.25\%. Being an optimization technique that gradually updates model parameters, SGD is helpful in handling large datasets. However, the performance of this model is somewhat disappointing in comparison to XGBoost and KNN—most probably due to its sensitivity to tuning hyperparameters, such as regularization and learning rate. Nevertheless, it is still a good representative of DDoS detection techniques because of its speed and scalability features, as in \cite{lam2020anomaly, khan2023neural}.

\vspace{\baselineskip}
\subsection{Naïve Bayes}
Relatively speaking, Naïve Bayes was the worst among the competitors, with an accuracy of 91.09\%, precision of 90.70\%, recall of 91.10\%, and F1 score of 90.90\%. The probabilistic nature of Naïve Bayes relies on the assumption of independence between features, which limits its ability to capture complex relationships in network traffic patterns. Since network traffic characteristics in IoT contexts can be interdependent, this deficiency can lead to misclassifications. Naïve Bayes has a low computational cost; thus, despite these limitations, it can still be applied in situations where speed and minimal resource usage are necessary \cite{pokhrel2021botnet, mirkovic2004taxonomy}.
\vspace{\baselineskip}
Table \ref{table:performance} provides the summarized results.

\begin{table}[htbp]
\caption{Performance Metrics of Machine Learning Models}
\begin{center}
\begin{tabular}{|c|c|c|c|c|}
\hline
Model & Accuracy & Precision & Recall & F1 Score \\
\hline
XGBoost & 99.82\% & 99.80\% & 99.85\% & 99.82\% \\
KNN & 99.56\% & 99.50\% & 99.60\% & 99.55\% \\
SGD & 98.25\% & 98.10\% & 98.30\% & 98.25\% \\
Naïve Bayes & 91.09\% & 90.70\% & 91.10\% & 90.90\% \\
\hline
\end{tabular}
\label{table:performance}
\end{center}
\end{table}

\begin{figure}[htbp]
\centerline{\includegraphics[width=\linewidth]{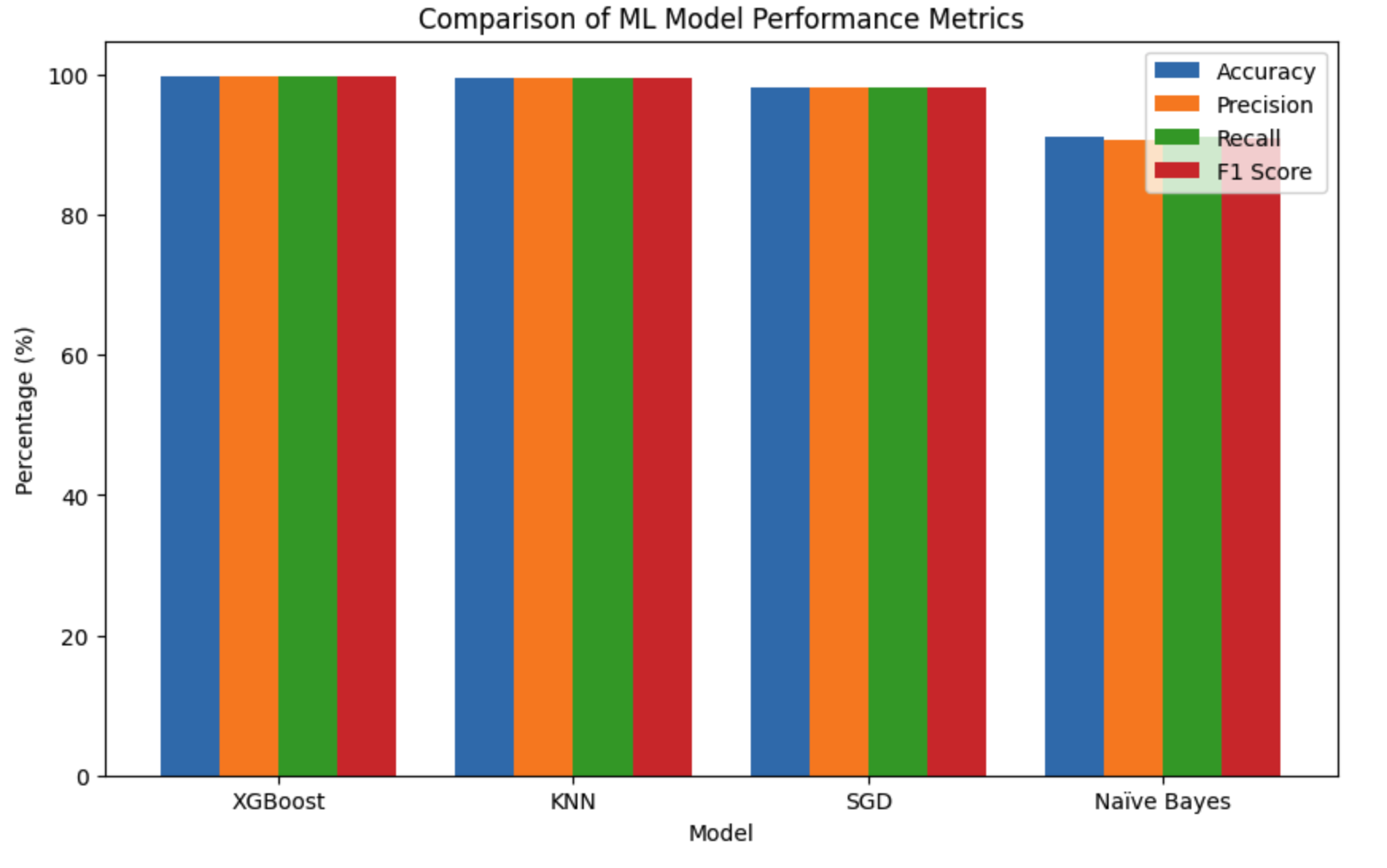}}
\caption{Performance Comparison of ML Models}
\label{fig:results_chart}
\end{figure}
\vspace{\baselineskip}
\section{Discussion}
XGBoost can be considered the most reliable model for the detection of DDoS attacks in IoT networks because it has shown the highest result in accuracy, precision, recall, and F1 score in comparison with all others. Although very good, both KNN and SGD are falling behind. XGBoost stands at the very top within the industry. Though Naive Bayes does really well in many tasks, it falters a little bit due to its lower recall and accuracy that resulted in misclassifying such assaults more often.
\vspace{\baselineskip}
\section{Conclusion}
Defences must advance along with the frequency and sophistication of DDoS assaults on IoT networks. This study will demonstrate how well several machine-learning models detect DDoS attacks; XGBoost performs the best across all assessed parameters, including accuracy, precision, recall, and F1 score. Reduction through regularization brings out the strength of XGBoost to overcome overfitting, while the handling of the interaction of features is sophisticated and hence suitable for complex, dynamic network traffic data. All these results showcase the big potential of machine learning in enhancing IoT security frameworks with dependable detection capabilities, important in preserving network integrity and availability.
\vspace{\baselineskip}
Naïve Bayes performed very poorly, while both K-Nearest Neighbours and Stochastic Gradient Descent had very good performances. That proves once again the inability of models predicated on the concept of probabilistic independence to handle the complexity introduced by IoT network traffic. This work emphasizes the importance of the choice of an appropriate machine learning model, fitted for the characteristics of IoT networks, where high accuracy, flexibility, and resilience against novel attack patterns play an important role.
\vspace{\baselineskip}
\section{Future Work}
Building upon the results of this study, future work can focus on several promising directions to enhance DDoS detection in IoT environments.
\vspace{\baselineskip}
\subsection{Model Optimization and Hybrid Approaches}
Future research can also be done in hybrid models, which combine the strengths of different algorithms. For example, combining Naïve Bayes' computational efficiency with XGBoost's precision may result in a fast and accurate model; real-time DDoS detection for IoT would be applicable in this context. Further, advanced feature engineering could optimize model performance by the selection and weighting of features according to their correlation with attack patterns.
\vspace{\baselineskip}
\subsection{Deep Learning Techniques}
The deep learning approach may include CNNs and RNNs in the analysis of network data with complex patterns, especially those showing sequential patterns that could indicate multi-vector DDoS attacks. Further, it could explore the transfer learning possibility where a model pre-trained on a similar dataset is adapted to the target dataset; this would further improve the training efficiency and also increase the accuracy of detection in an environment with a limited amount of labeled data.
\vspace{\baselineskip}
\subsection{Real-World Implementation and Testing}
Running such models in the context of real IoT infrastructures for real-time testing will deliver insights on their practical effectiveness. It will also be important to test the scalability of these models over IoT devices with limited resources, since IoT devices are generally lower in computational power compared to traditional network devices.
\vspace{\baselineskip}
\subsection{Ethical and Regulatory Considerations}
Ethical and regulatory considerations are very important while discussing the deployments of machine learning models in IoT networks. Privacy issues obviously come into consideration while dealing with huge volumes of network data, so privacy-preserving ML techniques are what researchers should look for in future works. Furthermore, the bias of ML algorithms should be addressed and mitigated to make the security applications free from discrimination.
\vspace{\baselineskip}
\section*{Acknowledgment}
I would like to thank Dr. Robert Abbas for his guidance and support throughout this research project. His expertise was invaluable in formulating the study’s objectives and evaluating its findings.
\vspace{\baselineskip}

\end{document}